\documentstyle[12pt,epsfig,epsf]{article}
\begin{document}

\newcommand{\nn}{\noindent}
\renewcommand{\thefootnote}{\fnsymbol{footnote}}
%%%%%% Begin   fey1
%Definition des diagrames de Feynman
% input of rpm.tex

%     un  Boson sortant:\unbs{}
\def\unbs#1{
\begin{picture}(100,200)(0,0)
\put(0,0){\circle*{5}}
\multiput(5,0)(8,-8){6}{\oval(8,8)[bl]}
\multiput(5,-8)(8,-8){6}{\oval(8,8)[tr]}
\put(49,-32){\makebox(0,0){#1}}
\end{picture}   }
%
%     Paire de  Bosons entrants:\be{}{}
\def\be#1#2{
\begin{picture}(100,200)(0,0)
\put(25,0){\circle*{5}}
\multiput(0,23)(8,-8){3}{\oval(8,8)[tr]}
\multiput(8,23)(8,-8){3}{\oval(8,8)[bl]}
\multiput(0,-23)(8,8){3}{\oval(8,8)[br]}
\multiput(8,-23)(8,8){3}{\oval(8,8)[tl]}
\put(-10,26){\makebox(0,0){#1}}
\put(-10,-26){\makebox(0,0){#2}}
\end{picture}   }
%   Paire de  Bosons sortants :\bs{}{}
\def\bs#1#2{
\begin{picture}(100,200)(0,0)
\put(0,0){\circle*{5}}
\multiput(5,0)(8,8){3}{\oval(8,8)[tl]}
\multiput(5,8)(8,8){3}{\oval(8,8)[br]}
\multiput(5,0)(8,-8){3}{\oval(8,8)[bl]}
\multiput(5,-8)(8,-8){3}{\oval(8,8)[tr]}
\put(34,30){\makebox(0,0){#1}}
\put(34,-30){\makebox(0,0){#2}}
\end{picture}   }
%  Paire de   Fermions entrants
\def\fe#1#2{
\begin{picture}(100,200)(0,0)
\put(34,0){\circle*{5}}
\put(0,35){\vector(1,-1){35}}
\put(-10,26){\makebox(0,0){#1}}
\put(-10,-26){\makebox(0,0){#2}}
\put(0,-35){\vector(1,1){35}}
\end{picture} }
%
%  Paire de   Fermions sortants
\def\fs#1#2{
\begin{picture}(100,200)(0,0)
\put(0,0){\circle*{5}}
\put(0,0){\vector(1,1){25}}
\put(34,26){\makebox(0,0){#1}}
\put(34,-26){\makebox(0,0){#2}}
\put(0,0){\vector(1,-1){25}}
\end{picture} }
% Boson propagateur :\bp{}{} propa:#2=24,entrant:#2=0,sortant#2=40
\def\bp#1#2{
\begin{picture}(100,200)(0,0)
\multiput(0,0)(16,0){3}{\oval(8,8)[t]}
\multiput(8,0)(16,0){3}{\oval(8,8)[b]}
\put(#2,15){\makebox(0,0){#1}}
\end{picture}   }
%
% Boson propagateur small :\sbp{}{} propa:#2=24,entrant:#2=0,sortant#2=40
\def\sbp#1#2{
\begin{picture}(100,200)(0,0)
\multiput(0,0)(16,0){2}{\oval(8,8)[t]}
\multiput(8,0)(16,0){2}{\oval(8,8)[b]}
\put(#2,15){\makebox(0,0){#1}}
\end{picture}   }
%
% Boson propagateur BIG :\bbp{}{} propa:#2=24,entrant:#2=0,sortant#2=40
\def\bbp#1#2{
\begin{picture}(100,200)(0,0)
\multiput(0,0)(16,0){5}{\oval(8,8)[t]}
\multiput(8,0)(16,0){5}{\oval(8,8)[b]}
\put(#2,15){\makebox(0,0){#1}}
\end{picture}   }
%
% Fermion propagateur :\fp{}{} propa:#2=24,entrant:#2=0,sortant#2=40
\def\fp#1#2{
\begin{picture}(100,200)(0,0)
\put(0,0){\line(1,0){48}}
\put(#2,15){\makebox(0,0){#1}}
\end{picture}   }
%
% Externel Fermion  :\fpe{}{} propa:#2=24,entrant:#2=0,sortant#2=40
\def\fpe#1#2{
\begin{picture}(100,200)(0,0)
\put(0,0){\line(1,0){35}}
\put(#2,15){\makebox(0,0){#1}}
\end{picture}   }
\def\fpp#1#2{
\begin{picture}(100,200)(0,0)
\put(0,0){\line(1,0){48}}
\put(#2,15){\makebox(0,0){#1}}
\end{picture}   }
%       Fermion voie t      :\ft{}
\def\ft#1{
\begin{picture}(100,200)(0,0)
\put(0,0){\line(0,1){48}}
\put(10,24){\makebox(0,0){#1}}
\end{picture}   }
\def\fth#1{
\begin{picture}(100,200)(0,0)
\put(0,0){\line(0,1){48}}
\put(10,24){\makebox(0,0){#1}}
\end{picture}   }
%       Boson voie t :\bt{}{}
\def\bt#1{
\begin{picture}(100,200)(0,0)
\multiput(0,0)(0,16){3}{\oval(8,8)[tl]}
\multiput(0,8)(0,16){3}{\oval(8,8)[br]}
\multiput(0,8)(0,16){3}{\oval(8,8)[tr]}
\multiput(0,0)(0,16){3}{\oval(8,8)[bl]}
\put(15,24){\makebox(0,0){#1}}
\end{picture}   }
\def\fpp#1#2{
\begin{picture}(100,200)(0,0)
\multiput(0,0)(10.8,0){6}{\line(1,0){5.2}}
\put(#2,15){\makebox(0,0){#1}}
\end{picture}   }
\def\hpe#1#2{
\begin{picture}(100,200)(0,0)
\multiput(0,0)(10.8,0){4}{\line(1,0){5.2}}
\put(#2,15){\makebox(0,0){#1}}
\end{picture}   }
\def\ftp#1{
\begin{picture}(100,200)(0,0)
\multiput(0,0)(0,10.8){5}{\line(0,1){5.2}}
\put(10,24){\makebox(0,0){#1}}
\end{picture}   }
%paire de bosons scalaires sortants
\def\fsp#1#2{
\begin{picture}(100,200)(0,0)
\put(0,0){\circle*{5}}
\multiput(0,0)(16,16){3}{\line(1,1){10}}
\multiput(0,0)(16,-16){3}{\line(1,-1){10}}
\put(27,36){\makebox(0,0){#1}}
\put(52,-36){\makebox(0,0){#2}}
\end{picture}   }
\def\fep#1#2{
\begin{picture}(100,200)(0,0)
\put(40,0){\circle*{5}}
\multiput(0,-40)(16,16){3}{\line(1,1){5}}
\multiput(0,40)(16,-16){3}{\line(1,-1){5}}
 \put(-7,36){\makebox(0,0){#1}}
\put(-7,-36){\makebox(0,0){#2}}
\end{picture}   }
%%%
% un boson scalaire sortants
\def\unfsp#1{
\begin{picture}(100,200)(0,0)
\put(0,0){\circle*{5}}
\multiput(0,0)(16,16){3}{\line(1,1){10}}
\put(45,36){\makebox(0,0){#1}}
\end{picture}   }
%%%%
%%%    end Fey1

\begin{flushright}
KEK Preprint 99/53\\
LPM/99/31\\
UFR-HEP/99/17 \\
July 1999 \\
\end{flushright}

\vspace*{.4cm}
\begin{center}
{\Large{ {\bf CP--odd Higgs boson production in association with}}}
{\Large{ {\bf Neutral gauge boson in High-Energy  $e^+e^-$ Collisions}}}

\vspace{.4cm}

{\large A.G. Akeroyd}$^{\mbox{1}}$,
{\large A. Arhrib}$^{\mbox{2,3}}$,
{\large M. Capdequi Peyran\`ere}$^{\mbox{4}}$

\vspace{.6cm}
{\sl
1: KEK Theory Group, Tsukuba\\
Ibaraki 305-0801, Japan

\vspace{.6cm}
2: D\'epartement de Math\'ematiques, Facult\'e des Sciences et Techniques\\
B.P 416, Tanger, Morocco

\vspace{.6cm}
3: UFR--High Energy Physics, Physics Department, Faculty of Sciences\\
PO Box 1014, Rabat--Morocco

\vspace{.6cm}
4: Laboratoire de Physique Math\'ematique et Th\'eorique, CNRS-UMR 5825\\
Universit\'e Montpellier II, F--34095 Montpellier Cedex 5, France
}
\end{center}

\setcounter{footnote}{4}
\vspace{1.2cm}

\begin{abstract}
\nn We study the associated production of a CP--odd
Higgs boson $A^0$ with a
neutral gauge boson (Z or photon) in high--energy $e^+ e^-$
collisions at the one--loop level in the framework of 
Two Higgs Doublet Models (THDM). We find that in the small $\tan \beta$
regime the top quark loop contribution is enhanced leading to 
significant cross--sections 
(about a few fb), while in the large 
$\tan \beta$ regime the cross--section does not attain observable 
rates.
\end{abstract}

\newpage

\pagestyle{plain}
\renewcommand{\thefootnote}{\arabic{footnote} }
\setcounter{footnote}{0}

\section*{1.~Introduction}
The discovery of a Higgs boson \cite{Higgs} is one of the major goals of
the present searches in particle physics. The Higgs boson of the
Standard Model (SM) is until now undiscovered but direct or indirect
results give stringent lower bounds on its mass. Moreover, the
problematic scalar sector of the SM can be enlarged
and some simple extensions of the SM with
several Higgs bosons have been intensively studied for many years.
The simplest extension is the Two Higgs Doublet
Model (THDM) \cite{gun}, the two most commonly studied versions being
classified as type I and II. They differ in how the Higgs
bosons are coupled to the fermions, although both versions possess
identical particle spectra after
electroweak symmetry breaking \cite{ewsb}.
From the 8 degrees of freedom initially present in the
2 Higgs doublets, 3 correspond to masses of the longitudinal gauge bosons,
leaving 5 degrees of freedom which should be manifested as 5
physical Higgs particles (2 charged Higgs $H^\pm$, 2 CP--even $H^0$, $h^0$
and one CP--odd $A^0$). Model type II is the structure found in the Minimal
Supersymmetric Standard Model (MSSM).
Until now no Higgs boson has been discovered, and from the null searches
one can derive direct and indirect bounds on their masses. The latest
such limits are $M_{H^{\pm}}> 69$ GeV \cite{ZHH} and $M_h + M_A>90$ GeV
\cite{ZAh}\footnote{There are still some regions which allow
$50$ GeV $\le M_h + M_A \le 90$ GeV.}.
We note that in the general THDM which will be considered in this
paper the existence of a very light $h^0$ or $A^0$ is not excluded by
the current searches (\cite{Kraw} and refs therein).

Higgs boson production in association with gauge bosons has been
extensively studied in the literature:
\begin{itemize}
\item Higgsstrahlung ($e^+e^- \to Z \Phi $)
 both for the SM Higgs boson \cite{HZ} or for  Higgs bosons
originating from extended Higgs sectors \cite{2HZ} (THDM and MSSM).
\item Associated production with a photon ($e^+e^-\to \gamma H^0/h^0/A^0 $)
\cite{abdelhak}, \cite{abdel}
\item Associated production of charged Higgs bosons
with a W gauge boson both at $e^+e^-$ colliders \cite{chinese}, 
\cite{achm},
and at Hadron colliders \cite{wh}.
\end{itemize}
CP--odd Higgs bosons can be produced at $e^+e^-$ colliders \cite{eecolliders}
via $e^+e^- \to h^0 A^0$ and $e^+e^- \to b\overline bA^0,t\overline tA^0$
\cite{eeA},\cite{Dawson}.
Pair production $e^+e^- \to A^0A^0$ is small \cite{PairHiggs}, as is the
analogous process at $\gamma\gamma$ colliders \cite{yyAA}.
In this paper we consider the rare decay $e^+e^-\to \gamma^*,Z^*\to ZA^0$
in the context of the THDM. Such a process is forbidden at tree--level
and proceeds via loops, with the rate at hadron colliders evaluated in
Ref. \cite{Kao}. Calculations of the loop mediated decay 
$A^0\to ZZ$ have appeared
in Refs. \cite{Mendez}, \cite{AZZ}, but not for the full process
$e^+e^-\to \gamma^*,Z^*\to ZA^0$.
Although not expected to give a rate comparable
to the mechanisms above, if observable this decay would
be a test of the THDM at the
one--loop level, since this mechanism would be sensitive to new physics
entering in the loop. In addition it is important to
have an accurate prediction
for $e^+e^-\to ZA^0$ in order to
distinguish between possible
signals from a CP--violating and CP--conserving THDM.
In a CP--violating THDM tree--level mixing is
allowed between the pure CP--even states
and pure CP--odd state, giving rise to 3 neutral Higgs scalars ($h_i$)
with no definite CP quantum numbers. Thus all three neutral scalars
may be produced in the Higgsstrahlung process
$e^+e^-\to Z^*\to Zh_i$, via a tree--level vertex 
$ZZh_i$. Therefore it is important to know
the possible magnitude of  $e^+e^-\to ZA^0$ in the CP--conserving THDM
in order to be sure if a (say) detected boson in the
Higgsstrahlung channel with a
low rate could be consistent with $A^0$, or is
actually evidence for a $h_i$ from a CP--violating THDM
with a small component of CP--even scalar field.
Such scalar--pseudoscalar 
mixing may also arise in the MSSM, being generated radiatively
if one allows CP--violating phases in the model \cite{CP}.
The paper is organized as follows. In section 2, we review all the
CP--odd Higgs boson interactions (with gauge bosons
 and fermions), section 3 contains the notation,
conventions and one--loop calculations while in section
4 we present our numerical results.
Finally section 5 contains our conclusions.

\renewcommand{\theequation}{2.\arabic{equation}}
\setcounter{equation}{0}
\section*{2. Notation, relevant couplings and One--Loop structure}
\subsubsection*{2.1 Notation and One--Loop structure}
We will use the following notation and conventions.
The momenta of the incoming electron and positron, outgoing
gauge boson $V$ and outgoing CP--odd Higgs boson
$A^0$ are denoted by $p_{e^-}$, $p_{e^+}$, $p_V$ and $p_{A^0}$,
respectively. Neglecting the electron mass $m_e$,
the momenta in the center of mass of the $e^+e^-$ system are given by:
\begin{eqnarray}
& & p_{e^-,e^+}=\frac{\sqrt{s}}{2} (1,0,0,\pm 1) \nonumber \\
& & p_{V,A^0}=\frac{\sqrt{s}}{2} (1\pm \frac{m_V^2 -M_{A}^2}{s},
\pm \kappa \sin\theta,0,\pm \kappa
\cos\theta), \nonumber
\end{eqnarray}
where $\sqrt{s}$ denotes the center of mass energy,
$\theta$ the scattering angle
between $e^+$ and $A^0$ and
$$\kappa ^2=(s-(M_{A} + m_V)^2)(s-(M_{A} -m_V)^2)/s^2 ;$$
$m_{V}$  and  $M_{A}$ are the masses  of the neutral gauge boson
 and of the CP--odd Higgs boson.\\
The Mandelstam variables are defined as follows:
\begin{eqnarray}
& & s =  (p_{e^-}+p_{e^+})^2 = (p_V+p_{A^0})^2  \nonumber\\
& & t = (p_{e^-}-p_V)^2 = (p_{e^+}-p_{A^0})^2 =
\frac{1}{2}(m_V^2  + M_{A}^2) -
\frac{s}{2}
+\frac{s}{2} \kappa \cos\theta  \nonumber\\
& & u = (p_{e^-}-p_{A^0})^2 = (p_{e^+}-p_V)^2 =
\frac{1}{2} (m_V^2 + M_{A}^2) -
\frac{s}{2}-
\frac{s}{2} \kappa \cos\theta \nonumber \\
& & s+t+u = m_V^2 + M_{A}^2  \nonumber
\end{eqnarray}

At the one--loop order (Fig.1.a--d),
the differential cross--section reads:
\begin{eqnarray}
\frac{d \sigma}{d \Omega}(e^+ e^- \to V A^0)= \frac{\kappa}{256 \pi^2 s}
\sum_{Pol}
\vert {\cal M}^1 \vert^2 .
\end{eqnarray}
The one--loop amplitude ${\cal M}^1 $ can be written as a sum of vertex
  and box contributions as follows:
$$ {\cal M}^1 = {\cal M}_V^{\gamma} + {\cal M}_V^{Z} +{\cal M}_B^{h} +
{\cal M}_B^{H} $$
where ${\cal M}_V^{\gamma}$ (${\cal M}_V^{Z}$)
denotes the vertex correction s--channel photon exchange
(s--channel Z exchange), depicted in Fig.1.a + Fig.1.b,
 and ${\cal M}_B^{h, H}$
denote the box contribution with $h^0$ and $H^0$ exchange (Fig.1.c).
All these contributions
fully project onto six invariants as follows:
$$ {\cal M}^1 =\sum_{i=1}^{6} {\cal M}_i {\cal A}_i$$
where the invariants ${\cal A}_i$ are given by:
\begin{eqnarray}
& & {\cal A}_1 = \bar{v}(p_{e^+}) \not\epsilon (p_V) \frac{1+\gamma_5}{2}
u(p_{e^-})\nonumber \\ & &
{\cal A}_2 = \bar{v}(p_{e^+}) \not\epsilon (p_V) \frac{1-\gamma_5}{2}
u(p_{e^-})\nonumber \\ & &
{\cal A}_3 =\bar{v}(p_{e^+}) \not p_V \frac{1+\gamma_5}{2} u(p_{e^-})
(p_{e^-}\epsilon(p_V))\nonumber \\ & &
{\cal A}_4=\bar{v}(p_{e^+}) \not p_V
\frac{1-\gamma_5}{2} u(p_{e^-}) (p_{e^-}\epsilon
(p_V))\nonumber \\ & &
{\cal A}_5=\bar{v}(p_{e^+}) \not p_V
\frac{1+\gamma_5}{2} u(p_{e^-}) (p_{e^+}\epsilon
(p_V))\nonumber \\ & &
{\cal A}_6=\bar{v}(p_{e^+}) \not p_V
\frac{1-\gamma_5}{2} u(p_{e^-}) (p_{e^+}\epsilon
(p_V)) \label{inva}
\end{eqnarray}
with $\epsilon$  the polarization of
the vector boson $V$.
Summing over the $V=\gamma, Z$ gauge boson polarizations,
the squared amplitude may be written as:\\
{\bf{Case V=Z}}
\begin{eqnarray}
& & \sum_{Z Pol} \vert {\cal M}^1 \vert^2 =
2 s (\vert {{\cal M}_1 \vert }^2 + \vert{{\cal M}_2 \vert }^2) -
\frac{(m_{A}^2m_Z^2 - t u)}{4 m_Z^2}
    \{ 4  \vert {{\cal M}_1 \vert }^2 + 4 \vert {{\cal M}_2 \vert }^2
\nonumber \\ & &  + 4 (m_Z^2 - t) Re[
{{\cal M}_1}{{\cal M}^*_3} +  {{\cal M}_2}{{\cal M}^*_4}]  +
 (m_Z^2 - t )^2 [ \vert  {{\cal M}_3 \vert }^2 +  \vert {{\cal M}_4}
\vert ^2]  \nonumber  \\
[0.2cm] & &
 +   4 (m_Z^2 - u) Re[ {{\cal M}_1} {{\cal M}^*_5} +
{{\cal M}_2} {{\cal M}^*_6}]
+  (m_Z^2 - u)^2 [  \vert {{\cal M}_5 \vert }^2 +
\vert {{\cal M}_6 \vert }^2]
\nonumber \\ [0.2cm] & &
 - 2 (m_{A}^2 m_Z^2 + m_Z^2 s - t u) Re[ {{\cal M}_3}  {{\cal M}^*_5} +
 {{\cal M}_4}  {{\cal M}^*_6} ]  \}
\end{eqnarray}
{\bf{Case V=$\gamma$}}
\begin{eqnarray}
 \sum_{\gamma\ Pol} \vert {\cal M}^1 \vert^2 & = &
     2s (  \vert {{\cal M}_1 \vert }^2  + \vert {{\cal M}_2 \vert }^2
) + s t (Re [ {\cal M}_1^{*} {\cal M}_3  +
{\cal M}_2^{*} {\cal M}_4 ]) +\nonumber \\ & &
      s u (Re [  {\cal M}_1^{*} {\cal M}_5  + {\cal M}_2^{*} {\cal M}_6  ])
-s t u (Re[ {\cal M}_3^{*} {\cal M}_5  + {\cal M}_4^{*} {\cal M}_6 ]  )
\end{eqnarray}

\subsection*{2.2 Relevant couplings for our study}
\subsubsection*{CP--odd Higgs boson interaction with fermions}
In the two Higgs doublet extensions of the Standard Model, there are
different ways of coupling the Higgs fields to matter. The most popular
are labelled as model type 
I and model type II. In the former, the quarks and leptons
couple only to the second Higgs doublet $\Phi_2$, 
while in the latter, in order to avoid
the problem of Flavor Changing Neutral Current (FCNC)
\cite{glashow-weinberg}, one
assumes that $\Phi_1$ couples only to down quarks (and charged leptons)
and $\Phi_2$ couples only to up quarks (and neutral leptons).
The type II model is the pattern found in the MSSM.

In general, the CP--odd Higgs interaction with 
fermions is given by:
\begin{eqnarray}
& & A^0 u\bar{u}=  Y_{uu} \gamma_5 \qquad , \qquad
A^0 d\bar{d}=  Y_{dd}  \gamma_5
\end{eqnarray}
Here $u$($d$) may refer to
any generation of up quark (down quark or charged leptons)
 and the $Y$ couplings are defined as follows:
\begin{eqnarray}
& & Y_{uu}= -\frac{gm_u}{2 M_W\tan\beta}\ \ \  \quad , \quad \ \ \
Y_{dd}= \frac{gm_d}{2  M_W \tan\beta} \ \ \ \ \ \quad 
\mbox{ Model I}    \nonumber \\
& & Y_{uu}= -\frac{gm_u}{2M_W\tan\beta} \quad \ \ \  , \ \ \  \quad
Y_{dd}= -\frac{gm_d\tan\beta }{2  M_W}\quad \ \ \ 
\mbox{ Model II}
\end{eqnarray}
It is worth noting the models I and II are not very 
different for the top--bottom loop corrections at low 
$\tan\beta$ because the term $m_t/\tan\beta$ will 
dominate and it is common to both types. 
In the case of large $\tan\beta$ the effects of down 
quarks and tau--lepton are enhanced (suppressed) 
in Model type II (Model type I).

\subsubsection*{2.3 Higgs boson interaction with gauge bosons}
In a general two Higgs doublet model, the interactions
between  Higgs bosons and gauge bosons are completely dictated by
the local gauge invariances and so they are
model independent regardless of whether the model is supersymmetric 
or not. These
interactions originate from the square of the covariant derivative
in the Higgs lagrangian which is given by:
\begin{eqnarray}
\sum_i (D_{\mu}\Phi_i)^+(D_{\mu}\Phi_i)=\sum_i [(\partial_\mu +ig \vec{T_a}
\vec{W_{\mu}^a}  +ig'\frac{Y_{\Phi_i}}{2}B_\mu) \Phi_i]^+(\partial_\mu +ig
\vec{T_a} \vec{W_{\mu}^a} +ig'\frac{Y_{\Phi_i}}{2}B_\mu )\Phi_i
\label{covder}
\end{eqnarray}
where: $\vec{T_a}$ is the isospin operator, $Y_{\Phi_i}$ the hypercharge of the
 Higgs fields,  $W^a_{\mu}$ the $SU(2)_L$ gauge fields,
$B_\mu$ the $U(1)_Y$ gauge field, and $g$ and $g'$ are the associated coupling
constants.\\
After expanding eq. (\ref{covder}) and deriving all the interaction vertices,
one finds, in accordance with the conservation of electromagnetic
current, that the $\gamma$--$A^0$--$Z$ vertex vanishes at
tree--level while the vertex $Z$--$Z$--$A^0$  vanishes by virtue of
CP--invariance \cite{323,324}.\\\
Therefore, at tree--level, the contributions  to
$e^+e^- \to Z A^0$ and/or $e^+e^- \to \gamma A^0$
only come from the $t$--channel via electron
exchange and from the following $s$--channels:
$e^+e^- \to h^* \to  Z A^0$ ($e^+e^- \to H^*\to Z A^0$)
where $H^0$ and $h^0$ are the two CP--even neutral Higgs fields.
All these contributions are
proportional to the electron mass $m_e$ which we will neglect in our work.
As a major consequence, in this limit, the studied process will be
only due to loop effects.\\
\\ 
From eq.(\ref{covder}) one can deduce the following vertices which
are needed for our study:
\begin{eqnarray}
& & Z_{\mu} Z_{\nu} h^0 =
\frac{g m_Z}{c_W} \sin{(\beta-\alpha)}g_{\mu\nu}  \ \qquad ,
\qquad \quad  Z_\mu Z_\nu H^0 =
\frac{g m_Z}{c_W} \cos{(\beta-\alpha)}g_{\mu\nu}  \nonumber \\
& & Z_{\mu} A^0 h^0 =\frac{g\cos{(\beta-\alpha)}}{2 c_W}
(p_h - p_A)_\mu \ , \
 Z_{\mu} A^0 H^0 = \frac{-g\sin{(\beta-\alpha)}}{2 c_W}(p_H - p_A)_{\mu}
\end{eqnarray}
where $c_W$ and $s_W$ are the cosine and sine of the Weinberg angle
 $\theta_W$.
We need also:
\begin{eqnarray}
& & Z_\mu q\bar{q} =
\gamma_\mu (g_q^L \frac{1-\gamma_5}{2} +
g_q^R \frac{1+\gamma_5}{2}) \ \ , \ \
Z_\mu e\bar{e} =
\gamma_\mu (g_V- g_A\gamma_5)\nonumber \\ 
& & \gamma_\mu q\bar{q} =
\gamma_\mu (-e_q) \ \ , \ \
\gamma_\mu e\bar{e} =
\gamma_\mu (-1)
\end{eqnarray}
where $q$ may refer to $u$ or $d$, and
\begin{eqnarray}
 & & g_{V}=(1-4s_W^2)/(4 s_W c_W)\ \ ,
 \quad  g_A = 1/(4 s_Wc_W) \nonumber \\
& &  g_{u}^{L}= -\frac{(1- 2 s_W^2 e_u)}{4 s_W c_W} \quad , \quad
  g_{d}^{L}= \frac{(1+ 2 s_W^2 e_d)}{4 s_W c_W} \qquad  , \quad
g_{u,d}^{R}= \frac{2s_W^2 e_{u,d}}{4 s_W c_W} \nonumber
\end{eqnarray}

\subsubsection*{2.4 CP--Violating scenario}
In a THDM, CP--violation (explicit or spontaneous) 
is possible in the Higgs sector if one
allows the discrete symmetry to be broken softly by a term
of dimension 2 ($\mu^2_{12}\Phi_1^{\dagger} \Phi_2 + h.c$).
In such a case mixing is permitted between
the pure CP--even states $h^0$, $H^0$ and the pure CP--odd state $A^0$,
leading to three neutral Higgs mass eigenstates which cannot be
assigned a definite CP quantum number.
In the notation of Ref.\cite{CPvHigg} these are referred
to as $h_1$, $h_2$, $h_3$, whose couplings obey various 
sum rules \cite{Mendez},\cite{Sum}
which in the CP--conserving case reduce to the familiar
sum rules of the THDM/MSSM e.g. one has (where the couplings 
are normalized to SM Higgs boson strength)
\begin{equation}
C_i^2+C^2_j+C^2_{ij}=1;\;\;\; C_1^2+C_2^2+C^2_3=1
\end{equation}
Here $i,j$ run from 1 to 3, $C_i$ is the coupling $ZZh_i$, and
$C_{ij}$ is the coupling $Zh_ih_j$.
Ref. \cite{Mendez} explained how CP--violation may be probed in the
Higgs sector by measuring non--zero values for each of $Zh_1h_2$, 
$Zh_2h_3$ and $Zh_3h_1$, or for each of
$Zh_ih_j$, $ZZh_i$,$ZZh_j$ (where $i\ne j$). 
A third way would be a positive signal
in all three Higgsstrahlung channels $e^+e^-\to h_1Z,h_2Z,h_3Z$, 
and it is possible that
the Higgs masses are arranged such that the latter production
mechanisms would be all open kinematically (i.e. $\sqrt s > m_Z + m_{h_i}$) 
while pair production of Higgs bosons are not.
Therefore the CP--violating THDM may provide measurable signals in the
$e^+e^-\to h_iZ$ channel for all three neutral Higgs bosons.
Hence in order to distinguish between a CP--violating THDM and a 
CP--conserving 
THDM in the above channels it is important
to know the attainable value of $e^+e^-\to ZA^0$ in the CP--conserving case.

In Refs. \cite{Barger},\cite{Hagiwara}, using effective lagrangians,
 it was shown that
a CP--odd Higgs scalar with effective point
like couplings $ZZA^0$ and $Z\gamma A^0$ would give rise to angular
distributions in the channel $e^+e^-\to \gamma^*,Z^*\to ZA^0$ which would
differ significantly from those in $e^+e^-\to Z^*\to Zh(H)$. 
Our analysis essentially determines the possible magnitude of these 
effective couplings
$ZZA^0$ and $Z\gamma A^0$ in the context of the THDM. In the notation of
\cite{Hagiwara} this would correspond to evaluating the arbitrary
couplings $\tilde b_{\gamma},\tilde b_{Z}$. Given the expected 
smallness of the rate for
$e^+e^- \to A^0 Z$, any angular distribution analysis for this channel 
would most likely be hampered by small statistics. If 
in the CP--violating THDM all three $ZZh_i$ couplings have a reasonable 
strength (i.e. significant pseudoscalar--scalar mixing), a detectable 
signal would
be possible which could not be mimicked by the CP--conserving THDM. In addition
there would be sufficient events to show that the angular distribution
proceeded via the CP--even scalar component.
The problematic case of interest to us is
when a $h_i$ is dominantly pseudoscalar and thus has a smaller (but still
detectable) rate 
in the $e^+e^- \to h_iZ$ channel. In this case the signal from 
$e^+e^- \to A^0Z$ is background to any possible interpretation of a 
CP--violating signal in this channel. It is also this scenario when angular
distributions would be affected by low statistics.

\renewcommand{\theequation}{4.\arabic{equation}}
\setcounter{equation}{0}
\section*{3. One-Loop Corrections}
We have evaluated the  pure radiative effects of the rare decay
$e^+ e^- \to  Z A^0$ ($e^+ e^- \to  \gamma A^0$ )
 at the one--loop level in the 't Hooft - Feynman gauge.
The sum of all the one--loop effects are ultra--violet (UV)
convergent but since some Feynman diagrams are UV divergent
we will use the dimensional
regularization scheme \cite{thooft} to deal with them.

The typical Feynman diagrams for the virtual corrections
of order $\alpha^2$ are drawn in figure 1. In the THDM,
the $\gamma-Z-A^0$ and the $Z-Z-A^0$ vertices receive corrections
only from fermion exchanges (Fig.1-a-b). There are also box diagrams
and their crossed analogies (fig.1-c).
Note that there is no mixing coming from the $Z-A^0$ 
($Z-G^0$) s--channel self--energy
because the vertices $\gamma-A^0-A^0$ ($\gamma-G^0-A^0$)
and $Z-A^0-A^0$ ($Z-G^0-A^0$) vanish, while the mixing
$Z-A^0$ has to be considered in the t--channel (fig.1.d). 
Owing to Lorentz invariance the $Z-A^0$
self energy is proportional to
$p_{A^0}^{\mu}=(p_{e^+}+p_{e^-}-p_V)^{\mu}$; then, since the vector boson 
V=$\gamma$ or Z is on--shell,
the t--channel amplitude will be
proportional to $m_e$ and consequently vanishes. Note also that the 
tadpole diagrams have vanishing contributions.

In the on--shell scheme defined in \cite{Dabelstein-Hollik},
it is found that  
there is no counter--term for the Z--Z--$A^0$, Z--$\gamma$--$A^0$ and  
$\gamma$--$\gamma$--$A^0$ vertices. 
Consequently the one--loop vertices
Z--Z--$A^0$, Z--$\gamma$--$A^0$   and $\gamma$--$\gamma$--$A^0$ 
have to be separately UV finite  
and this feature will provide us with a good check of our calculations.

All the Feynman diagrams are generated and
computed using FeynArts
\cite{seep} and
FeynCalc \cite{rolf} packages supplied with some Mathematica "know-how".
We also use  the fortran FF--package \cite{ff} in the numerical analysis.
\\
\\
{\bf{Vertex correction: Fig.1.a + Fig.1.b: $e^+e^-\to A^0 Z$}}\\
\\
In terms of the  $B_0$, $C_0$ and $D_i$ Passarino--Veltman functions,
 ${\cal M}_V^{\gamma}$ and ${\cal M}_V^{Z}$ are given by\footnote{Note that 
the one--loop vertex $A^0\to ZZ$ has been evaluated in \cite{Mendez}.} :
\begin{eqnarray}
{\cal M}_V^{\gamma} &=& 2 N_C e_q m_q Y_{qq} \frac{\alpha^2}{s}
\{ 2( {\cal A}_3 -  {\cal A}_4 -  {\cal A}_5 +  {\cal A}_6) +
     (t-u)(   {\cal A}_1 - {\cal A}_2 ) \} ({g_{q}^L} + g_{q}^R)
\nonumber \\ & &
      C_0(M_A^2,m_Z^2, s, m_q^2, m_q^2, m_q^2)\\
{\cal M}_V^{Z} &=&\frac{-2N_C \alpha^2 m_q Y_{qq}}{(s-m_Z^2)\kappa^2}
\{ (g_A - g_V) (2 {\cal A}_3 - 2 {\cal A}_5 + (t-u) {\cal A}_1 )\nonumber \\
&  + & (g_A + g_V) (2 {\cal A}_4 -
 2 {\cal A}_6   + {\cal A}_2 (t - u)\}
 (  ({{g_{q}^L}} - g_{q}^R)^2 [
2 M_A^2 B_0(M_A^2, m_q^2, m_q^2)\nonumber \\ &  - &
        (M_A^2 + m_Z^2 - s)B_0(m_Z^2, m_q^2, m_q^2)  - 
        (M_A^2 - m_Z^2 + s) 
  B_0(s, m_q^2, m_q^2)]\nonumber \\ &  + &
  ( [ M_A^2 m_Z^2  - m_Z^4 +
M_A^2 s  +  2 m_Z^2 s - s^2 ] ({g_q^L}^2 + {g_{q}^R}^2) \nonumber \\
 & +  & 2 [- M_A^4  + M_A^2 m_Z^2 + M_A^2 s] g_q^L g_q^R )
         C_0(M_A^2, m_Z^2, s, m_q^2, m_q^2, m_q^2))
\end{eqnarray}
Here $e_q$ and $m_q$ denote the charge and
the mass of the specific quark;
$g_V$, $g_A$, $g_q^L$ and $g_q^L$ are
the couplings defined before.\\
As defined in the appendix A, the $B_0$ functions are UV divergent whereas 
the $C_0$ are UV convergent. Consequently it is easy to check that
the one--loop vertices defined below are UV convergent, as they should be.
\\
\\
{\bf{Vertex correction: Fig.1.a + Fig.1.b: $e^+e^-\to A^0 \gamma$}}\\
\\
In a similar way we found:
\begin{eqnarray}
{\cal M}_V^{\gamma} &=& -4 N_C e_q^2 m_q Y_{qq} \frac{\alpha^2}{s}
\{ 2( {\cal A}_3 -  {\cal A}_4 -  {\cal A}_5 +  {\cal A}_6) +
     (t-u)(   {\cal A}_1  - {\cal A}_2 ) \} 
\nonumber \\ & &
      C_0(M_A^2,0, s, m_q^2, m_q^2, m_q^2)\\
{\cal M}_V^{Z} &=& 
-2N_C  e_q \frac{\alpha^2 m_q Y_{qq}}{s-m_Z^2} \{   (g_A + g_V)(2 {\cal A}_4 - 
              2 {\cal A}_6 + (t-u) {\cal A}_2 )\nonumber \\ & +  &    (g_A - 
g_V)(2 {\cal A}_3 -
             2 {\cal A}_5 +(t-u) {\cal A}_1 )  
  \}  ( {g_{q}^L} + g_{q}^R )
           C_0(M_A^2,0,s,m_q^2,m_q^2,m_q^2)
\end{eqnarray}
$\ $
\\
{\bf{Box corrections: Fig.1.c}}:\\
\\
Box corrections are only present for $e^+e^- \to A^0 Z$, 
because the photon does not couple to neutral particles.
\begin{eqnarray}
{\cal M}_B^{h} & = &\frac{\alpha^2 \sin{2(\beta-\alpha)} m_W}{4 s_W^2 c_W^3}
[ 4 {\cal A}_3 (g_{A} - g_V)^2 D_1 +
4 {\cal A}_4 (g_A + g_V)^2 D_1 -
{\cal A}_1 (g_A - g_V)^2 \{
\nonumber \\ & & C_0(m_e^2, m_e^2, s, m_Z^2, m_e^2, m_Z^2) +
      (M_h^2 - t) D_0 + (m_Z^2 - t) D_1 +
( t - M_A^2) D_3 \} - \nonumber \\ & &
{\cal A}_2 (g_A + g_V)^2 \{
 C_0(m_e^2, m_e^2, s, m_Z^2, m_e^2, m_Z^2) +
   (M_h^2 - t) D_0 + (m_Z^2 - t) D_1
\nonumber \\ & &  + (t - M_A^2)  D_3 \} ]
\end{eqnarray}
All the $D_i(i=0,1,2,3)$ have the same arguments:
$(m_Z^2, m_e^2, m_e^2, M_A^2, t, s, M_h^2, m_Z^2, m_e^2, m_Z^2)$.  
The amplitude of a crossed box can be obtained from the 
direct one by changing:
$t\to u $, ${\cal A}_3\to {\cal A}_5$,
${\cal A}_4\to {\cal A}_6$ and a global sign. Moreover
${\cal M}_B^{H}$ can be obtained from ${\cal M}_B^{h}$ just by
changing $M_h$ in $M_H$ and a global sign.

%PaVe[3, {MZ^2, 0, 0, MA^2, U, S},{Mh^2, MZ^2, 0, MZ^2}]
\vspace{-0.8cm}
\begin{picture}(100,100)(0,0)
\put(4,0) {\fe{$e^+$}{$e^-$}}
\put(41,0) {\bp{$\ \ \ \ \gamma , Z$}{8}}
\put(88,0){\line(1,1){28}}
\put(88,0){\line(1,-1){28}}
\put(117,-27){\line(0,1){51}}
\put(116,25){\circle*{7}}
\put(116,-25){\circle*{7}}
\put(88,0){\circle*{5}}
\put(92,16){\makebox{$q$}}
\put(89,-19){\makebox{$q$}}
\put(119,0){\makebox{$q$}}
\put(116,25) {\hpe{$\ $}{10}}
\put(129,29) {$A^0$}
\put(115,-25) {\bp{Z, $\gamma$}{30}}
\put(99,11){\vector(-1,-1){1}}
\put(117,0){\vector(0,1){1}}
\put(101,-13){\vector(1,-1){1}}
\put(55,-45){\makebox{Fig.1.a}}
\put(246,0) {\bp{$\ \ \ \ \ \gamma , Z$}{8}}
\put(295,0){\line(1,1){28}}
\put(295,0){\line(1,-1){28}}
\put(324,-27){\line(0,1){51}}
\put(323,25){\circle*{7}}
\put(323,-25){\circle*{7}}
\put(295,0){\circle*{5}}
\put(207,0) {\fe{$e^+$}{$e^-$}}
\put(299,16){\makebox{$q$}}
\put(296,-19){\makebox{$q$}}
\put(326,0){\makebox{$q$}}
\put(323,25) {\hpe{$\ $}{10}}
\put(339,28) {$A^0$}
\put(324,-25) {\bp{Z, $\gamma$}{30}}
\put(306.2,11){\vector(1,1){1}}
\put(324,0){\vector(0,-1){1}}
\put(308,-13){\vector(-1,1){1}}
\put(261,-45){\makebox{Fig.1.b}}
\end{picture}
$\ $
\vspace{1.cm}
\par
\begin{center}
\begin{picture}(100,100)(0,0)
\setlength{\unitlength}{1.pt}
\put(-60,50) {\fpe{$e^+$}{0}}
\put(-20,50){\circle*{5}}
\put(-20,48) {\bp{Z}{24}}
\put(29,48){\circle*{5}}
\put(-22,1) {\ft{$\ $}}
\put(-26,22) {$e$}
\put(-20,1){\circle*{5}}
\put(-19,0) {\bp{Z}{24}}
\put(-60,0) {\fpe{$e^-$}{0}}
\put(29,2){\circle*{5}}
\put(27,48) {\hpe{$\ $}{24}}
\put(55,53) {$A^0$}
\put(59,7) {$Z$}
\put(25,0) {\ftp{$\ \ \ h (H)$}}
\put(27,1) {\bp{$\ $}{30}}
\put(87,20){\makebox{\quad +\quad crossed diagram}}
\put(20,-24){\makebox{Fig.1.$c$}}
\end{picture}
\end{center}

\par
\begin{center}
\begin{picture}(100,100)(0,0)
\setlength{\unitlength}{1.pt}
\put(-60,50) {\fpe{$e^+$}{0}}
\put(-20,50){\circle*{5}}
\put(29,48){\circle{30}}
\put(-20,50) {\sbp{Z}{24}}
\put(13,50){\circle*{5}}
\put(-22,1) {\ft{$\ $}}
\put(-26,22) {$e$}
\put(-20,1){\circle*{5}}
\put(-60,0) {\fpe{$e^-$}{0}}
\put(42,48) {\hpe{$\ $}{20}}
\put(46,48){\circle*{5}}
\put(55,53) {$A^0$}
\put(59,7) {$\gamma , Z$}
\put(-19,1) {\bbp{$\ $}{30}}
\put(87,20){\makebox{\quad +\quad crossed diagram}}
\put(20,-24){\makebox{Fig.1.$d$}}
\put(20,-54){\makebox{{\bf{Figure.1}}}}
\end{picture}
\end{center}
$\ $
\\
\\
\renewcommand{\theequation}{4.\arabic{equation}}
\setcounter{equation}{0}
\section*{4. Numerical results and discussion}
In this section we focus on
the numerical analysis.
We take the following experimental input for the physical parameters
\cite{databooklet}:
\begin{itemize}
\item the fine structure constant: $\alpha=\frac{e^2}{4\pi}=1/137.03598$.
\item the gauge boson masses: $m_Z=91.187\ GeV$ and  $m_W=80.41\ GeV$.
In the on--shell scheme we consider, $\sin^2 \theta_W$ is given by
$\sin^2 \theta_W\equiv 1- \frac{m_W^2}{m_Z^2}$, which is not modified by loop 
corrections.
\item the top--bottom quark masses are taken to be: $m_t=175$ GeV and
$m_b=4.5$ GeV. The tau lepton mass is taken to be $m_\tau=1.77$ GeV
\end{itemize}
As pointed out in the introduction, in the general THDM 
the CP--odd Higgs mass
is not constrained by experiment, and only the sum $M_h +M_A$ is constrained 
to be heavier
than 90 GeV. Therefore in our analysis we are free to consider
 a very light CP--odd Higgs mass. Taking into account perturbative 
requirements on the Yukawa couplings $\mbox{tan}\beta$ 
is constrained to be in the range: $ 0.1 \leq  \mbox{tan} \beta \leq 80 $.

Let us first discuss briefly the THDM radiative contributions to the CP--odd 
$A^0$ associated production with a photon which has been considered before in 
\cite{abdel}. We stress here that we are in perfect agreement with \cite{abdel}
both on analytical result and numerical result.  We would like to highlight 
in Fig.2.a the enhancement of the cross--section for small $\tan\beta$. 
One can reach a cross--section of about 0.8 fb 
for $\mbox{tan}\beta=0.3$ (8 fb for 
$\mbox{tan}\beta=0.1$). 

Fig.2.b shows the total cross--section against $\tan\beta$ for CP--odd 
$A^0$ associated production with a Z gauge boson. One can see also the 
enhancement of the cross--section in the small $\tan\beta$  regime.
This enhancement for small $\mbox{tan}\beta$ is common to 
both Model~I and Model~II. The most important 
contribution in the fermionic case
comes from top quark loops. In the large $\tan\beta$ regime in Model~II 
the corrections start to 
be sensitive to the bottom, light quarks and tau lepton loops. 
As mentioned in section 3, in the case of associated 
CP--odd Higgs production with a Z gauge boson, there are extra 
box contributions to the cross--section. In all Figures 2a, 2b, 3a and 3b
we have chosen $M_h=95$ GeV, $M_H=180$ GeV and $\tan\alpha=+2.1$. 
It is found that box contributions to the cross 
section are rather small. 
For large $\tan\beta$ the fermionic contributions also become small and
of the same order of magnitude as the box contributions. 

In Fig.3.a we show the dependence of the cross--section
 on the CP--odd Higgs mass for $\sqrt{s}=500$ GeV 
in the case where $\tan\beta=0.2,0.5$ and $1.6$. 
One can see that the cross--section 
is maximized for both
low $\tan\beta$ and small $M_A$ mass. 
The kink in the figure
corresponds to $M_A>2m_t$ i.e. when the decay $A^0\to t\bar{t}$ opens.
At $\sqrt s=500$ GeV the cross--section for the SM Higgs boson in the 
Higgsstrahlung channel falls from $50$ fb to $20$ fb as the Higgs mass 
increases from 200 GeV to 300 GeV. Fig 3a shows that
for $\tan\beta=0.2$ one finds $\sigma(e^+e^-\to AZ)$ falling 
from 0.2 fb to 0.1 fb in
the same mass region. If one wished to interpret 
a third (by definition the smallest) Higgsstrahlung signal
as evidence for a CP--violating THDM, a 
cross--section comfortably in excess of 1 fb would be required.
We recall from eq. 2.10 that $\sum_{i=1}^3 C^2_i=1$ (normalized to
SM strength), and so 
$\sigma(e^+e^-\to AZ)$ constitutes a non--negligible background. 

In Fig.3.b we show the total cross--section  
as a function of the
center of mass energy $\sqrt{s}$ for $M_A=50,150, 270, 400 $ GeV and 
for $\tan\beta=0.5$. One can find a total
cross--section of about 0.04 fb only for the CP--odd Higgs mass in the 
intermediate range 
($M_A$=50--150 GeV) and for $\sqrt{s}\approx 500$ GeV.
At high center of mass energy the rate of production 
is rather small.

\renewcommand{\theequation}{5.\arabic{equation}}
\setcounter{equation}{0}
\section*{5. Summary}
To conclude, we have computed the associated production of 
a CP--odd Higgs boson
with a gauge boson $V=Z,\gamma$ in high energy $e^+e^-$ collisions
in the framework of the general two Higgs doublet model. 
The calculation is performed within the dimensional regularisation
scheme.

We have shown that in the small $\tan \beta$
regime the top quark loop contribution is enhanced leading to 
significant cross--sections 
(about a few fb), while in the large 
$\tan \beta$ regime the cross--section does not attain observable 
rates.

The smallness of the cross--section means that there is no chance
of seeing such associated production at LEP--II, while such 
production could be relevant for a Next Linear Collider machine and
would require a very high luminosity option.

\vspace{1.cm}

{\bf Acknowledgment:} 
A. Akeroyd
was supported by the Japan Society for Promotion of Science (JSPS).
A. Arhrib acknowledges the Abdus Salam International Centre for 
Theoretical Physics for the kind
hospitality during his visit where part of this work has been done.

\renewcommand{\theequation}{B.\arabic{equation}}
\setcounter{equation}{0}

\newpage
\renewcommand{\theequation}{A.\arabic{equation}}
\setcounter{equation}{0}

\subsection*{Appendix A: Passarino--Veltman Functions}
Let us briefly recall the definitions of scalar and tensor integrals we use.
The inverses of the propagators are denoted by
\begin{eqnarray} 
d_0= q^2-m_0^2 \ , \ d_i= (q+p_i)^2-m_i^2 \nonumber
\end{eqnarray}
where the $p_i$ are the momenta of the external particles.\\
\\
{\bf Two point function:}
\begin{eqnarray} 
 B_{0}(p_1^2,m_0^2,m_1^2)=\frac{(2\pi \mu)^{(4-D)} }{i \pi^2}\int 
d^Dq \frac{1}{d_0 d_1} 
\nonumber 
\end{eqnarray}
$\mu$ is an arbitrary renormalization scale and $D$ is the space--time 
dimension.\\
\\
{\bf Three point function:}
\begin{eqnarray} 
 C_{0} (p_1^2,p_{12}^2,p_2^2,m_0^2,m_1^2,m_2^2)=
\frac{(2\pi \mu)^{(4-D)} }{i \pi^2}\int d^Dq \frac{ 1}{d_0 d_1 d_2} \nonumber
\end{eqnarray}
where $p_{ij}^2=(p_i -p_j)^2$. 
\\
{\bf Four point functions:}
\begin{eqnarray}
D_{0,\mu} (p_1^2,p_{12}^2,p_{23}^2,p_3^2,p_2^2,p_{13}^2,m_0^2,
m_1^2,m_2^2,m_3^2)=
\frac{(2\pi \mu)^{(4-D)} }{i \pi^2}\int d^D q \frac{ 1, q_{\mu}}
{d_0 d_1 d_2 d_3} 
\end{eqnarray}
Using Lorentz invariance, one can write the vectorial function $D_\mu$ in 
terms of new
scalar functions $D_i$ as follows:
\begin{eqnarray}
& & D_\mu  =  {p_1}_{\mu} D_1 +  p_{2\mu} D_2 +  p_{3\mu} D_3 
\end{eqnarray}
Moreover one can always get all the "new" 
functions $D_i$ in terms of the fundamental ones, $A_0$, $B_0$, $C_0$ and 
$D_0$.
The analytical expressions of all the scalar functions can be found in 
Ref. \cite{pv}

\newpage

\subsection*{Figure Captions}

\vspace*{0.5cm}

\renewcommand{\labelenumi}{{\bf Fig.} \arabic{enumi}}
\begin{enumerate}

\item  %{\bf Fig. 1} 
One--Loop diagram contributions to  
$e^+e^- \to A^0 V$ ($V=\gamma$ or $Z$). 
Fig.1.a + Fig.1.b are fermionic vertex corrections. 
Fig.1.c are box diagrams and Fig.1.d is the t--channel self energy.

\item  %{\bf Fig. 2}
{\bf Fig.2.a}: Total cross--section for $e^+e^- \to A^0 \gamma $ 
(vertex only) as a function of 
$\tan\beta$
for $\sqrt{s}=500$ GeV and  several values of $M_A$. 
\\
\\
{\bf Fig.2.b}: Total cross--section for $e^+e^- \to A^0 Z $ (vertex + boxes) 
as a function of $\tan\beta$
for $\sqrt{s}=500$ GeV and several values of $M_A$. 

\vspace{5mm}
\item  %{\bf Fig. 3}
{\bf Fig.3.a}: Total cross--section for $e^+e^- \to A^0  Z $ (vertex + boxes)  asa function of $M_A$
for $\sqrt{s}=500$ GeV and  several values of $\tan\beta$. 
\\
\\
{\bf Fig.3.b}:Total cross--section for $e^+e^- \to A^0  Z $ (vertex + boxes) as
a function of $\sqrt{s}$
for $\tan\beta=0.5$  and for  several values of $M_A$.

\end{enumerate}

\newpage
\renewcommand{\thepage}{}
\begin{minipage}[t]{19.cm}
\setlength{\unitlength}{1.in}
\begin{picture}(1.9,1.9)(1.2,8.6)
\centerline{\epsffile{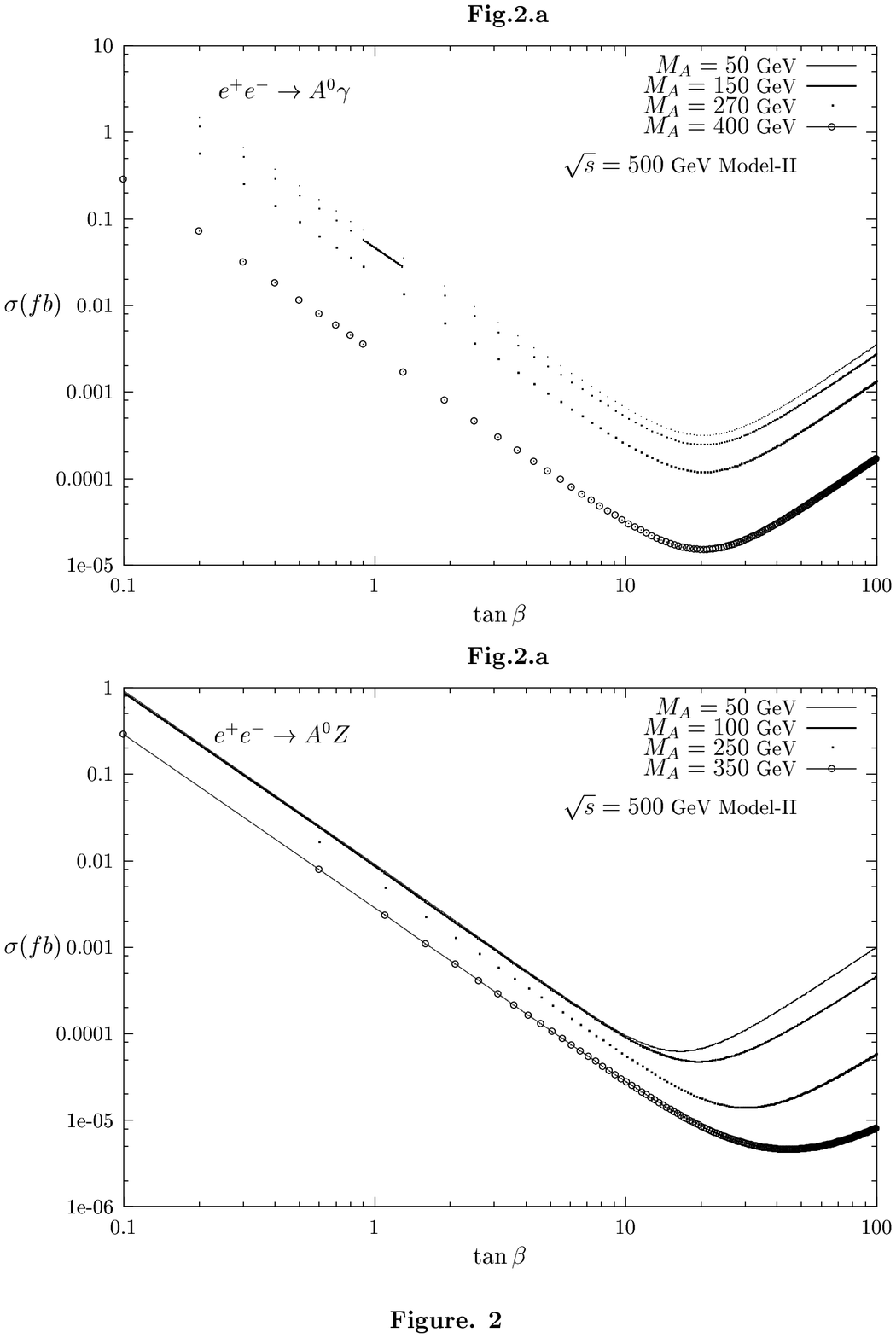}}
\end{picture}
\end{minipage}

\newpage
\begin{minipage}[t]{19.cm}
\setlength{\unitlength}{1.in}
\begin{picture}(1.9,1.9)(1.2,8.6)
\centerline{\epsffile{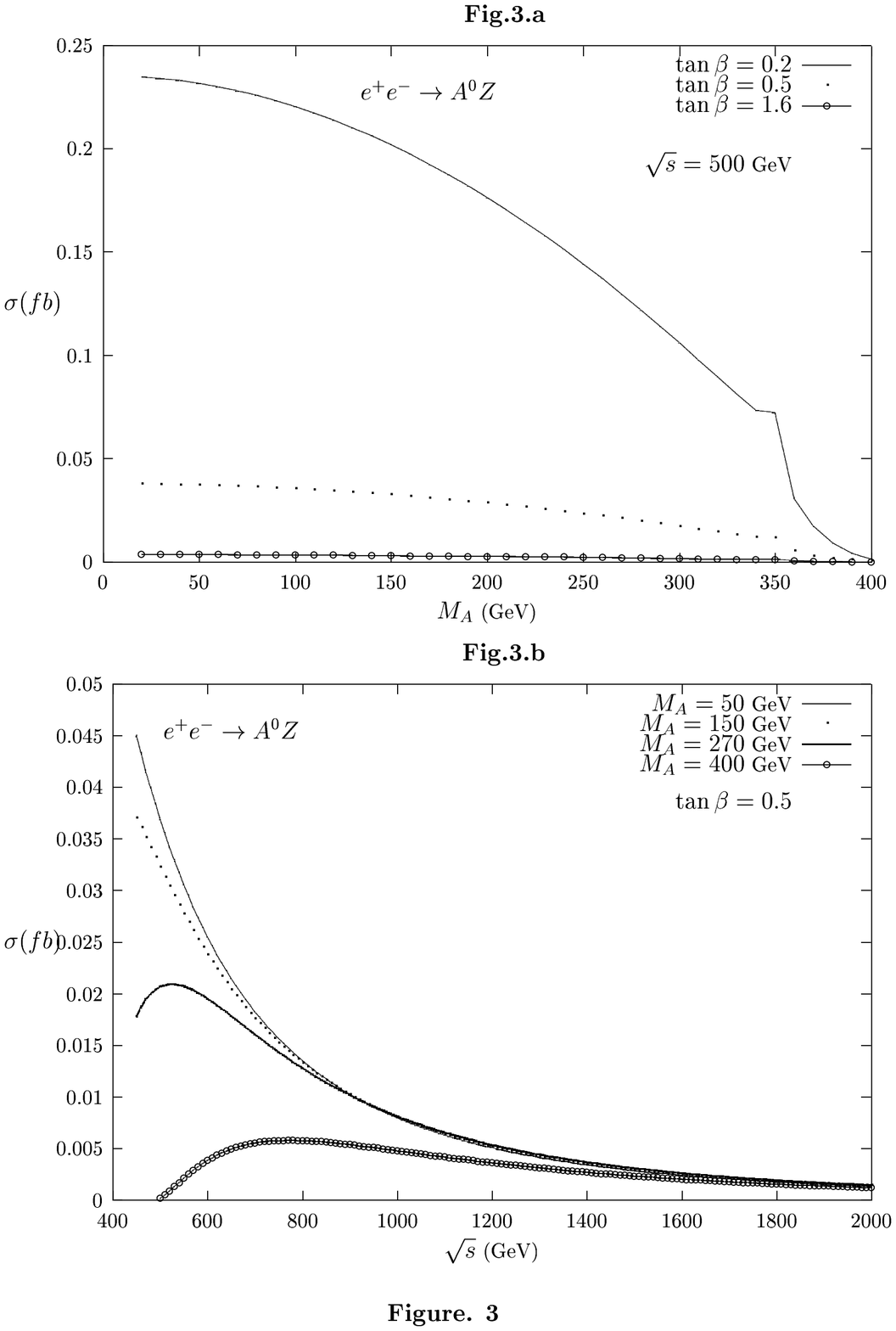}}
\end{picture}
\vspace{18.cm}

%\hspace{7.cm}{\bf Figure.3 }
\end{minipage}

\end{document}